\documentclass[namedreferences]{solarphysics}

\usepackage[optionalrh,solaromanenum]{spr-sola-addons} 

\usepackage{graphicx}        
\usepackage{color}           
\usepackage{hyperref}
\usepackage{bm}

\newcommand{\eq}[1] {Equation~\ref{#1}}

\newcommand{\aap}{    {\it Astron. Astrophys.}}

\newcommand{\aapr}{   {\it Astron. Astrophys. Rev.}}
\newcommand{\apss}{   {\it Astrophys. Space Sci.}}

\newcommand{\apj}{    {\it Astrophys. J.}}
\newcommand{\apjl}{   {\it Astrophys. J. Lett.}}

\newcommand{\jfm}{  {\it J. Fluid Mech.}} 
\newcommand{\jatms}{  {\it J. Atmos. Sci.}} 

\newcommand{\mnras}{  {\it Mon. Not. Roy. Astron. Soc.}}
\newcommand{\nat}{    {\it Nature}}

\newcommand{\pasj}{   {\it Pub. Astron. Soc. Japan}}

\newcommand{\solphys}{{\it Solar Phys.}}

\newcommand{\jpmg}{   {\it J. Phys. A. Math. Gen.}} 
\newcommand{\roysocproca}{ {\it Roy. Soc. London Proc. Ser. A}}
\newcommand{\lrsp}   { \it{Liv. Rev. Solar Phys.}}
\ifx \arxivurl  \undefined \def \arxivurl#1{\href{http://arxiv.org/abs/#1}{\textsf{arXiv}}}\fi
\ifx \urlurl  \undefined \def \urlurl#1{\href{#1}{\textsf{URL}}}\fi
\chardef\us=`\_

\begin{document}

\begin{article}
\begin{opening}

\title{Amplitudes of Solar Gravity Modes: A Review}

\author[addressref={aff1},corref,email={kevin.belkacem@observatoiredeparis.psl.eu}]{\inits{K.}\fnm{K.}~\lnm{Belkacem}\orcid{0000-0003-4548-4347}}
\author[addressref={aff2}]{\inits{C.}\fnm{C.}~\lnm{~Pin\c{c}on}\orcid{0000-0002-5490-7033}}
\author[addressref={aff3}]{\inits{G.}\fnm{G.}~\lnm{Buldgen}\orcid{0000-0001-6357-1992}}

\address[id=aff1]{LESIA, Observatoire de Paris, CNRS, Université PSL, Sorbonne Université, Université Paris Cité, 5 place Jules Janssen, 92195 Meudon, France}
\address[id=aff2]{LERMA, Observatoire de Paris, PSL University, CNRS, Sorbonne University, Paris, France}
\address[id=aff3]{Astronomy Department, University of Geneva, Versoix, 1290, Switzerland}

\runningauthor{K. Belkacem et al.}
\runningtitle{Amplitudes of Solar Gravity Modes}

\begin{abstract}
Solar gravity modes are considered as the {\it Rosetta Stone} for probing and subsequently deciphering the physical properties of the solar inner-most layers. Recent claims of positive detection therefore shed some new light on the long-standing issue of estimating solar gravity mode amplitudes. In this article, our objective is to review the theoretical efforts intended to predict solar gravity mode amplitudes. Because most of these studies assumed analogous driving and damping properties to those for the observed acoustic modes, we also provide a short overview of our current knowledge for these modes in the Sun and solar-type stars (which show solar-like oscillations) before diving into the specific problem of solar gravity modes. Finally, taking recent estimates into account, we conclude and confirm that the low-frequency domain (typically between $10\,\mu$Hz and $100\,\mu$Hz) is certainly more suited to focus on for detecting solar gravity modes. More precisely, around $60\,\mu$Hz, the theoretical estimates are slightly lower than the observational detection threshold as provided by the GOLF (Global Oscillations at Low Frequencies) instrument by about a factor of two only. This is typically within the current uncertainties associated with theoretical estimates and should motivate us for improving our knowledge on turbulence in the whole solar convective region, which is key for improving the accuracy of $g$-mode amplitude estimates. The recent detection of solar inertial modes (\citeauthor{Gizon2021}, \textit{Astron. Astrophys.} \textbf{652}, L6, \citeyear{Gizon2021}) combined with the continuous development of numerical simulations provide interesting prospects for future studies.   
\end{abstract}
\end{opening}

\section{Introduction}\label{s:intro}

Since the first detection of solar acoustic modes by \citet{Leighton1962} and \citet{Evans1962} and their interpretation by \cite{Ulrich70} and \cite{Leibacher71}, helioseismology has become a powerful way for probing the solar interior and testing the physics of solar models \citep[see][for a comprehensive review]{JCD2021}. However, we are still almost unable to probe the innermost layers of the Sun and this motivated decades of active search for solar gravity modes (hereafter $g$-modes). Indeed, their restoring force is buoyancy and as such they present large amplitudes in the radiative region while being evanescent in the convective region. Consequently, they are recognized as powerful probes of the solar radiative layers. Solar $g$-modes would be decisive for answering key questions in solar and stellar physics such as to identify and characterize the physical processes responsible for the redistribution of angular momentum in the radiative region and subsequently allowing us to reproduce the observed rotation profile in the Sun and solar-type stars, that is, which show solar-like oscillations. Three main candidates have been suggested to solve this issue: fossil magnetic fields \citep{Gough1998}, internal gravity waves \citep{Charbonnel2005}, and magnetic instabilities \citep{Eggenberger2005}, all of them predicting a different rotation of the solar core. By now, solar $p$-modes have only allowed to probe down to 0.2 solar radii, missing information on the inner core. As shown by \citet{Corbard1998}, completely different solutions can indeed be obtained depending on the dataset and the inversion techniques used, leading us to conclude that we do not have any reliable estimate of the rotation of the solar core yet. This emphasizes the importance of detecting $g$-modes to constrain the angular momentum transport in radiative regions. In addition, the detection of solar $g$-modes would offer crucial independent constraints on the solar-core conditions, allowing tests of the physics of nuclear reactions such as electronic screenings \citep{Bahcall2002,Shaviv2003,Mussack2011,Mussack2011ApSS,Salmon2021} and nuclear reaction rates, independently of neutrino-detection experiments.

In that context, $g$-modes were early considered worth deploying efforts for their detection. The first claims of detection of solar gravity modes started with the works of \cite{Severny76} and \citet{Brookes1976}. Unfortunately, none of them were confirmed even after more than ten years of observations from SOHO \citep{Appourchaux2000} and such a disappointing pattern repeated several times \citep[a comprehensive review of the detection attempts can be found in ][]{Appourchaux2010}. The most recent detection claims were made by \citet{Garcia07} and \citet{Fossat2017}. The former investigated the low-frequency domain, with the hope of detecting high radial-order $g$-modes. The method looked for regularities in the power spectrum, and the authors  claimed to detect a periodicity in accordance with what is expected from  simulated power spectra. Unfortunately, this detection was not confirmed \citep{Appourchaux2010}. The latest detection claim by \citet{Fossat2017} \citep[see also][]{Fossat2018}  indirectly found evidence for the signature of low-frequency gravity modes in the temporal variations of the large frequency separation of high-frequency acoustic modes. Their analysis led to the reported identification of hundreds of gravity modes, from which they could infer the asymptotic period spacing and the mean core rotation rate. Again, the detection has not been confirmed and was questioned by \citet{Schunker2018} and \citet{Appourchaux2019}. Almost in parallel, theoretical studies focused on the coupling between acoustic and gravity modes and reinforced the fragility of the detection claim \citep{Scherrer2019,Boning2019}. Therefore, the quest of solar $g$-modes, while at present unsuccessful, is still ongoing. 

In that spirit, analysis and observational efforts were (and still are) accompanied (and sometimes driven) by theoretical developments and calculations primarily intended to estimate $g$-mode amplitudes at the solar photosphere. They are also necessary to design future observational missions and guide future seismic studies. The first theoretical estimates mainly considered the Reynolds stress of turbulent convective eddies as the source mechanism \citep[e.g.][]{Gough85,Kumar96,Belkacem2009}. These estimates were heavily based on an analogy with observed solar acoustic modes because it was assumed that the driving and damping processes are the same. More precisely, it was considered that modes are excited by turbulent convection within the convective region while the damping is dominated by the convection--oscillation couplings in the same region. While such assumptions can be viewed as reasonable, two main difficulties arise for a quantitative estimate of $g$-mode amplitudes. First, the driving is expected to occur in the bulk of the solar convective region in which our knowledge of the properties of turbulence is quite limited. Second, our understanding of mode damping remains very approximate so that estimating $g$-mode damping from our knowledge of $p$-modes does not guarantee to grasp the main physics nor to provide reliable results. Note that, for solar $p$-modes, these two issues are mitigated by both the use of 3D numerical simulations which provide the properties of turbulent convection in the uppermost convective region, and direct observational constraints on mode damping based on the measurements of mode linewidths. As we will discuss in detail in the following sections, $g$-mode amplitude estimates based on analogies with observed $p$-modes must nevertheless be considered with caution. 

Other mechanisms were explored, such as excitation by magnetic torques \citep{Dziembowski1985} or mode coupling \citep{Dziembowski1983,Wolff2007}. However, among the candidates, penetrative convection at the base of the solar convective region is certainly a promising one and has been identified for a long time to be potentially able to drive progressive gravity waves \citep[e.g.][]{Schatzman1996} and gravity wave normal modes \citep[e.g.][]{Andersen96}. This assertion was mainly motivated by laboratory
experiments and theoretical studies for atmospheric flows \citep[e.g.][]{Townsend1966,Stull1976} and is now supported by numerical simulations. The main picture is the following; strong downdrafts originating from diving cool granules at the solar surface develop by turbulent entrainment of matter as coherent structures when crossing the convective region \citep[e.g.][]{Turner86,Rieutord95}. As these plumes reach the bottom of the convective zone, they can penetrate into the underlying stably stratified radiative layers; there, the plumes are braked by buoyancy and can transfer a part of their kinetic energy into gravity waves (either progressive waves or normal modes). This excitation mechanism is ubiquitous in numerical simulations of extended convective envelopes overlying radiative zones \citep[e.g.][]{Dintrans2005,Rogers05,Rogers2006b,Rogers2013,Alvan2014,Edelmann2019,Lesaux22} but the covered values of the physical parameters are far from stellar regimes (for instance the P\'eclet number of turbulent plumes, the Prandtl number, etc...). Hence, while numerical simulations are useful tools to gain some insight, quantitative estimates by means of semi-analytical excitation models are  required to allow for quantitative estimates \citep[e.g.][]{Pincon2021}. 

In this article, our objective is to review some of the theoretical efforts intended to predict solar gravity-mode amplitudes, and to that end  we split the subject into two main themes. First, in Sections~\ref{driving_turb} and \ref{mode_g_convection}, we focus on the works based on the assumption that driving and damping are governed by turbulent convection. Because most of these studies assumed analogous driving properties with observed modes, we begin with a short review of our current knowledge for these modes in the Sun and solar-type stars before diving into the specific problem of solar gravity modes. In that spirit, we propose an estimate of solar mixed-mode amplitudes based on an analogy with mixed modes in evolved stars. Second, in Section~\ref{penetrative_conv}, we consider $g$-mode driving by penetrative convection. Our lack of direct constraints on the penetrative region making the modeling of this driving process difficult to quantify, we briefly review how numerical simulations can be used to gain some physical insight in Section~\ref{help_simu3D} before discussing the recent estimates by semi-analytical models in Section~\ref{semi_analytics}. 

\section{Driving and Damping by Turbulent Convection}
\label{driving_turb}

Before addressing the issue of solar $g$-mode amplitudes, we will succinctly review our current understanding of the physical mechanisms responsible for mode driving and damping of observed $p$-modes in solar-type stars that show solar-like oscillations. This is motivated by $g$-mode amplitudes estimates based on analogies with $p$-modes,  i.e. under the assumption that the driving and damping are similarly due to turbulent convection. 

\subsection{Where Do We Stand for Solar-Like Oscillations?}\label{solarlike}

Since the discovery of solar oscillations much work has been undertaken to understand mode driving and damping. At the beginning, a stability analysis was performed by \citet{ando75}  and \citet{antia82} who concluded that, in the absence of a dynamic coupling between convection and oscillations, most solar acoustic modes are found to be unstable. In contrast, by including the effect of turbulent pressure or turbulent viscosity, solar modes were shown to be stable by \citet{gk77a} and \citet{Balmforth1992a}. As the quality of the solar observations improved, normal modes were shown to exhibit a Lorentzian profile in the power spectrum (which corresponds to an exponentially damped oscillation in the time domain), lifting the issue of their stability and elevating the issue of the physics behind mode driving and damping. 

For mode driving, however, the first studies began with the works by \citet{unno62} and \citet{stein67} who considered wave generation by turbulence by generalizing the approach of \citet{lighthill52} to a stratified atmosphere. They concluded that Reynolds stresses should be the main source of acoustic wave generation. The mode forcing was thus very quickly identified and, with the exception of a transient debate on the relative contributions of Reynolds stresses and the non-adiabatic contribution of the gas pressure fluctuations, this conclusion is still favoured nowadays \citep[see, for instance, the review by][]{Samadi2015}. In this framework, a notable leap forward has been made by \citet{gk77b}. Despite an underestimate of the amplitudes, pointed out by \citet{osaki90}, this work still constitutes the basis of the current formalisms for modelling the forcing of solar-type modes. Since this seminal work, different evolutions of this model have been developed by \citet{dolginov84,b92c,gk94,Houdek1999,samadi2001,chaplin05,samadi02II,Belkacem2006,Belkacem2008,Belkacem2010}, and these approaches differ from each other in the way they describe the turbulent convection. More recently, with the help of 3D hydrodynamical simulations of the uppermost convective region of solar-type stars, these formalisms have been successfully confronted by the observations \citep[see for instance the reviews by][]{Samadi2011,Samadi2015} and one can conclude that the main physics has been captured. Even the related and long-standing problem of mode asymmetries and the asymmetry reversal in intensity/velocity has recently been addressed with some success \citep{Philidet2020a,Philidet2020b}. Consequently, for solar-type stars and except for some remaining issues for red-giant stars, mode driving is now essentially understood all along their evolution \citep[see for instance][for a review]{Belkacem2013}.

For mode damping, the situation is more complex in the sense that there is no clear consensus on the dominant physical mechanisms at work. While the forcing is sensitive to the main properties of turbulent convection such as the profile of turbulent velocities and the eddy-time-correlation spectrum, the damping is sensitive to the interplay between oscillations and turbulent convection. Given that the thermal time-scale, the turbulent time-scale, and the modal periods are almost the same in the super-adiabatic layers, it is challenging to disentangle and identify the main physical mechanism responsible for mode damping. More precisely, two main contributions have been identified to play a crucial role among many others: the dynamic damping (related to the modulation of the turbulent pressure by the oscillation) and the thermal damping (related to the modulation of the non-adiabatic contribution of the gas pressure fluctuation). They are in fact of the same order of magnitude but of opposite sign and the sum of these contributions is a residual, hence the difficulty and the numerous debates on the subject \cite[see][for a review]{HoudekMAD2015}. It is therefore quite remarkable to note that this ``marginal stability" was brought to light very early by \citet{gk77a} and nicely illustrated by \citet{goldreich91} using simple order-of-magnitude arguments. Later works, which improved the physical description of the involved processes and in particular of turbulent convection, alternated in their conclusions as to which process was dominant for mode damping. \citet{Gough80} and \citet{b92c} found that damping is dominated by turbulent pressure modulation, while \citet{mad05} and \citet{mad06c} pointed out the dominant role of convective flow modulation and \citet{Belkacem2012} again pointed out the role of turbulent pressure modulation. Therefore, mode damping currently remains an open issue. To overcome these difficulties, going beyond the standard approaches based on the mixing-length theories is definitively needed. This currently motivates the direct investigation of the normal modes of 3D numerical simulations \citep[e.g.][]{Belkacem2019,Zhou2020} or the development of new theories such as recently proposed by \citet{Philidet2021a,Philidet2022}.

To summarize, while the physics of mode driving is rather well understood and modeled, mode damping remains the main source of uncertainties. When dealing with $g$-mode amplitude estimates in Section~\ref{mode_g_convection}, this state-of-the-art is to be recalled for assessing their validity and for appreciating the adopted strategies.

\subsection{The Special Case of Mixed Modes in Evolved Low-Mass Stars}
\label{driving_turb_mixed}

The physics of mode driving and damping recently received a renewed interest with the detection of mixed modes in evolved low-mass stars. Because in Section~\ref{analogy_g} we will propose an estimate of solar mixed modes based on an analogy with mixed modes in evolved stars, it is worth introducing and summarizing the recent main findings regarding their driving and damping. 

  \begin{figure}    
   \centerline{\includegraphics[width=0.9\textwidth,clip=]{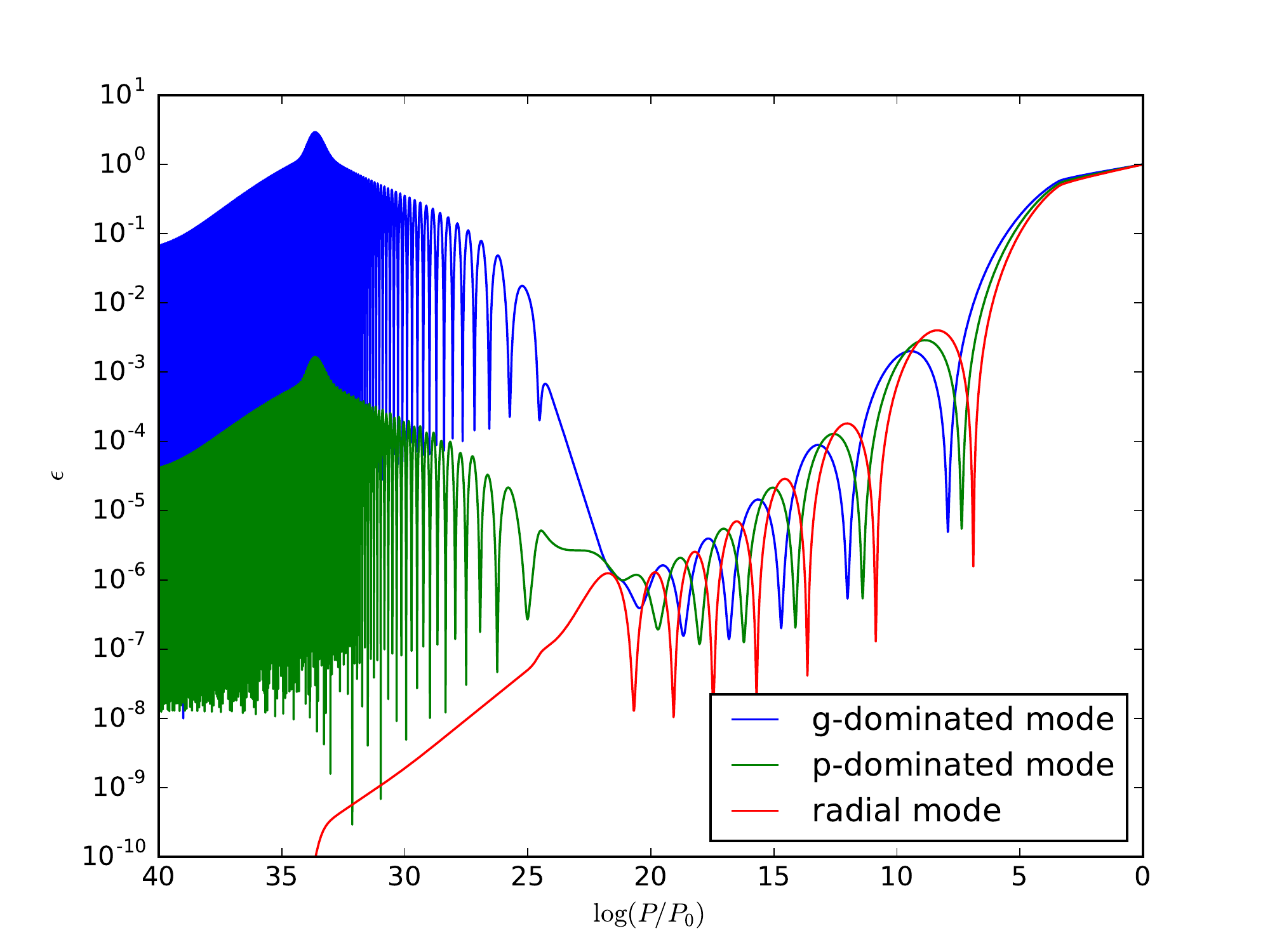}}
              \caption{Normalized kinetic energy for a radial mode (\textit{red line}), a $p$-dominated mode (\textit{green line}), and a $g$-dominated mode (blue line) versus the logarithm of the pressure ($P_0$ is the pressure at the surface). The equilibrium model corresponds to an evolved red-giant (RGB) star \citep[see][for details of the model]{Belkacem2015a}.}
   \label{plot_energies}
   \end{figure}
   
  \begin{figure}    
   \centerline{\includegraphics[width=0.9\textwidth,clip=]{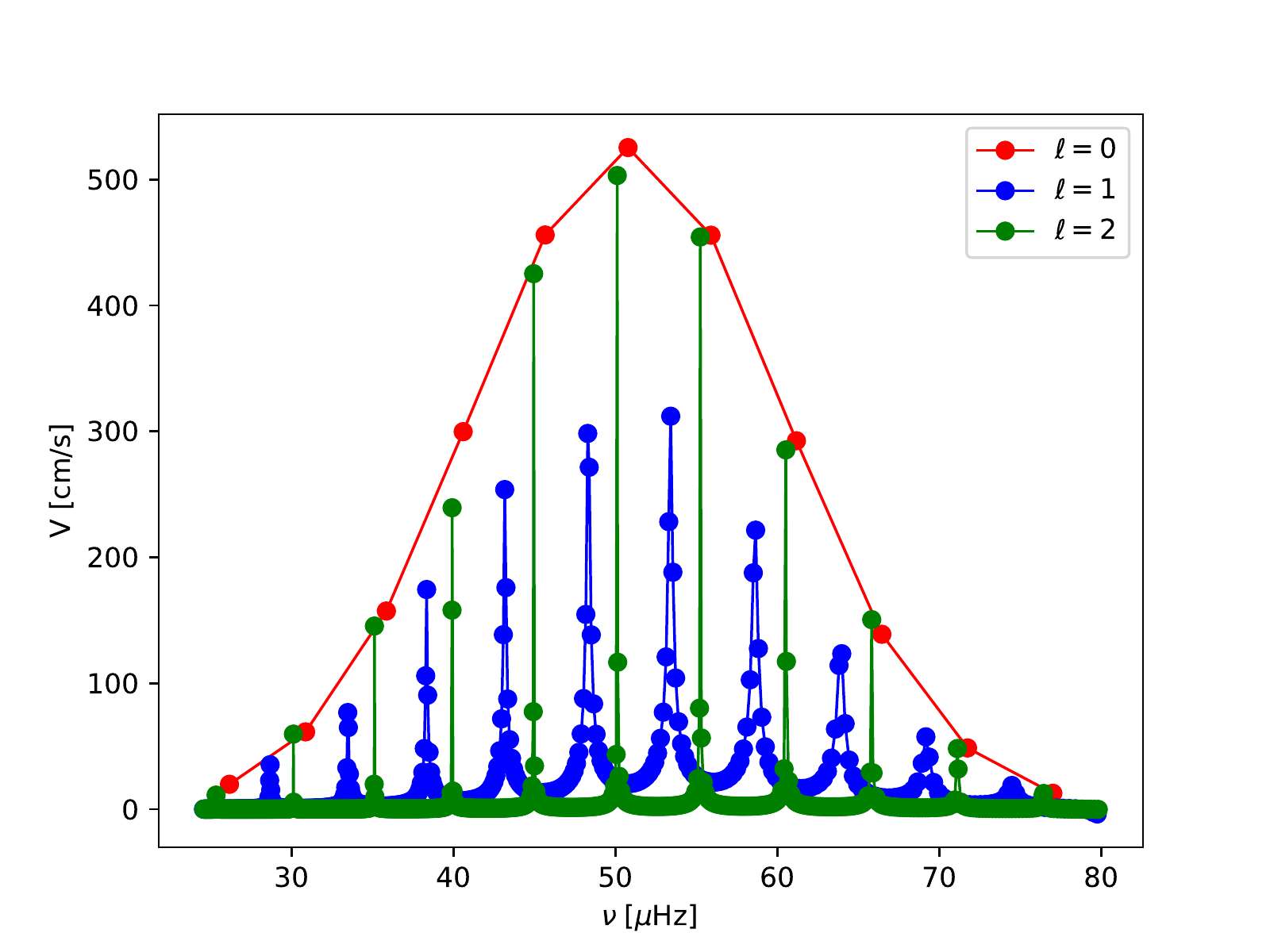}}
              \caption{Mode amplitudes versus mode frequencies computed as described by \citet[][]{Belkacem2015a} for an evolved red-giant  star.}
   \label{plot_amplitudes}
   \end{figure}

As a star evolves off the main sequence, its core contracts and its envelope expands so that the frequency range of $g$-modes increases (due to an increase of the buoyancy frequency in the radiative interior) and the frequency range of the $p$-modes decreases (due to the decrease of the surface gravity and so the frequency of the maximum height in the power spectrum). There is then an overlap that enables the establishment of mixed modes \citep[][]{Dziembowski2001,Dupret2009}, which show high amplitudes both in the inner radiative region and in the outer convective region (see Figure~\ref{plot_energies}). They can thus be easily detectable while bringing information on the inner-most layers of the stars. These mixed modes, as named in the early works of \citet{Dziembowski1971} for Cepheids and \citet{Scuflaire1974} for a condensed polytropic model, have been subject to an extensive investigation from a theoretical point of view \citep[e.g.][]{Shibahashi1979,Dziembowski2001,JCD2004}. 
Their detection by CoRoT (Convection, Rotation and planetary Transits) and {\it Kepler} enabled a rich harvest of physical information on the internal structure and evolution all along the evolution of low-mass stars \citep[see for instance the reviews by][]{Chaplin2013,Mosser2016,Hekker2017,Belkacem2019b}. 

Mixed modes have been extensively investigated to understand what governs their amplitudes \citep[e.g.][]{Dupret2009,Grosjean2014} and it has been shown that, compared to a $p$-mode of similar frequency, mainly two physical ingredients are susceptible to affect them, namely  the mode trapping and an extra contribution of mode damping in the radiative region. Actually, the dominant contribution of the latter is what we call the {\it radiative damping} since normal modes with a non-negligible amplitude lose energy radiatively. The influence of radiative damping has already been extensively investigated for other classes of pulsators \cite[see][for details]{Samadi2015} and, for  red giants, its influence has been investigated from a theoretical point of view by \cite{Dziembowski2001,Dupret2009,Dziembowski2012,Grosjean2014} to estimate the evolutionary status for which the damping will suppress (or at least limit) mixed-mode amplitudes. The main conclusion is that radiative damping becomes dominant either for high-angular degrees or highly evolved red-giant stars. The former effect is mode trapping because, if one compares the mixed mode amplitude to a radial mode of similar frequency, the amplitude (this holds for both mode driving and damping) will be modulated by mode inertia. Indeed, if a non-radial mode is efficiently trapped in the core, its (surface) amplitude will be diminished compared to a radial mode. The frequency dependence of the surface amplitude (see Figure~\ref{plot_amplitudes}) is therefore related to the trapping efficiency or in other words to the relative amplitude of the eigenfunctions in the core and envelope of the star. The effect of the trapping can be so important around a given radial mode in red-giant stars that it can lead to the non-detection of close non-radial mixed modes.

\section{The Solar Gravity Modes: Driving by Turbulent Convection}
\label{mode_g_convection}

We will now dive into the specific case of solar gravity modes and more precisely into their excitation and damping by turbulent convection.

\subsection{Analogies with \textit{p}-mode driving and damping}
\label{analogy_p}

Most of the $g$-mode amplitude estimates have been performed based on an analogy with observed $p$-modes. Indeed, these determinations assumed that either the driving or the damping experienced by gravity modes are of the same origin, although there are some subtle differences. 

In this framework, one of the first quantitative estimates of $g$-mode amplitudes was performed by \citet{Gough85} based on an equipartition assumption. It consists in equating the mode energy with the kinetic energy of resonant eddies whose lifetimes are close to the modal period. This approach presents the major advantage of by-passing the computation of both excitation and damping processes. 
 \citet{Gough85} found a maximum of velocity of about $0.5 \, \textrm{mm\,s}^{-1}$ for the dipolar modes near $\nu \approx 100\,\mu$Hz while quadripolar modes are found to reach values around $5\,\textrm{mm\,s}^{-1}$ near $\nu \approx 400\,-\,500\,\mu$Hz (see Figure~\ref{bilan_amplitudes}). To assess these estimates, it is fundamental to figure out what are the assumptions behind the equipartition hypothesis. For solar $p$-modes, this assumption has been theoretically justified by \citet{gk77b} assuming that the modes are damped by turbulent stresses \citep[modelled using an eddy viscosity as initially proposed by][]{Ledoux58} and excited by Reynolds stresses. The main weakness of the approach is therefore the way mode damping is modeled. Indeed, as mentioned in Section~\ref{solarlike}, mode damping is now understood to be the residual between two main contributions (dynamic and thermal dampings) while the equipartition assumption only considers dynamic damping using a very crude approximation. It is therefore unlikely that the equipartition assumption relies on a solid theoretical foundation, making the estimates of $g$-mode amplitudes by \citet{Gough85} debatable and certainly leading to an overestimate of amplitudes (see Figure~\ref{bilan_amplitudes}). 
 
Going a step further, \citet{Kumar96} proposed an estimate of $g$-mode amplitudes based on the full computation of both excitation and damping rates. It has been performed using the \cite{gk94} formalism for the driving (calibrated on the observed $p$-mode excitation rates) in which the authors assumed  that the kinetic energy of the eddies scale according to the Kolmogorov spectrum. More importantly,  \citet{Kumar96} assumed turbulent eddies as temporally correlated following a Gaussian spectrum. For the damping, both turbulent and radiative contributions were considered, as proposed by \cite{goldreich91}, and it was found that $g$-modes are long-lived modes with life times ranging from $10^6 \, \textrm{years}$ for high frequencies to $10^3 \, \textrm{years}$ for higher-order modes. Subsequently, the surface velocities were found to be (for  $\ell=1$ modes) about $0.3\,\textrm{mm}\,\textrm{s}^{-1}$ near $\nu = 300\,\mu\textrm{Hz}$  and $0.02\,\textrm{mm}\,\textrm{s}^{-1}$ for $\nu < 100\,\mu\textrm{Hz}$. In retrospect, these estimates suffer from two main deficiencies. First, the way the eddies are time-correlated now clearly appears to be inadequate. As shown by \citet{Belkacem2010}, to reproduce the excitation rates of solar $p$-modes, the eddy-time-correlation function must be modeled using (in the Fourier domain) a Lorentzian function for turbulent frequencies below the inverse of the Eulerian micro-scale (it is the temporal equivalent of the Taylor micro-scale, which corresponds to the largest scale at which viscosity affects the dynamic of eddies) and a Gaussian function for higher frequencies. Second, the computation of the damping rates as proposed by \citet{Kumar96} still remains quite approximate (except for high-order $g$-modes for which the radiative damping is dominant and rather well modelled) and suffers from our ignorance and inability to properly model them for solar $p$-modes. 

Guided by these two limitations, \citet{Belkacem2009} proposed to reassess $g$-mode amplitudes. The formalism used by \cite{Belkacem2009} to compute excitation rates of non-radial modes was  developed by \cite{Belkacem2008}, who extended to non-radial modes the work of \cite{samadi2001}. Compared to the work of \citet{Kumar96}, the excitation rates are modelled using a Lorentzian eddy-time correlation function using 3D numerical simulations from the ASH code \citep{Miesch2008}, which permits us to get constraints on this physical ingredient within the whole convective region. Damping rates were computed with the fully non-radial non-adiabatic pulsation code MAD \citep{mad05,mad06c} by taking into account the variations of the convective flux, the turbulent pressure and the dissipation rate of turbulent kinetic energy. It was found that above $110\,\mu\textrm{Hz}$ the damping rates are sensitive to the convection/pulsation interactions in the uppermost layers of the solar convective regions. Therefore, they restricted the $g$-mode amplitude estimates to lower frequencies for which the damping is essentially radiative and can be estimated in a reliable way. Taking  visibility factors as well as the limb-darkening into account, \citet{Belkacem2009} finally found that the maximum of apparent surface velocities of asymptotic $g$-modes is $\approx 3$ mm\,s$^{-1}$ for $\ell=1$ at $\nu\approx 60\,\mu$Hz and $\ell=2$ at $\nu\approx 100\,\mu$Hz. Those results then put the theoretical $g$-mode amplitudes near the GOLF observational threshold, but one must keep in mind that these estimates rely on turbulent properties and in particular velocities obtained using a numerical simulation, which does not guarantee a perfect estimate even if the degree of realism is higher than mixing-length estimates. 

\subsection{Analogy with Mixed Modes in Evolved Low-Mass Stars}
\label{analogy_g}

\begin{figure}[!ht]
	\begin{center}
		\includegraphics[width=0.9\textwidth,clip=]{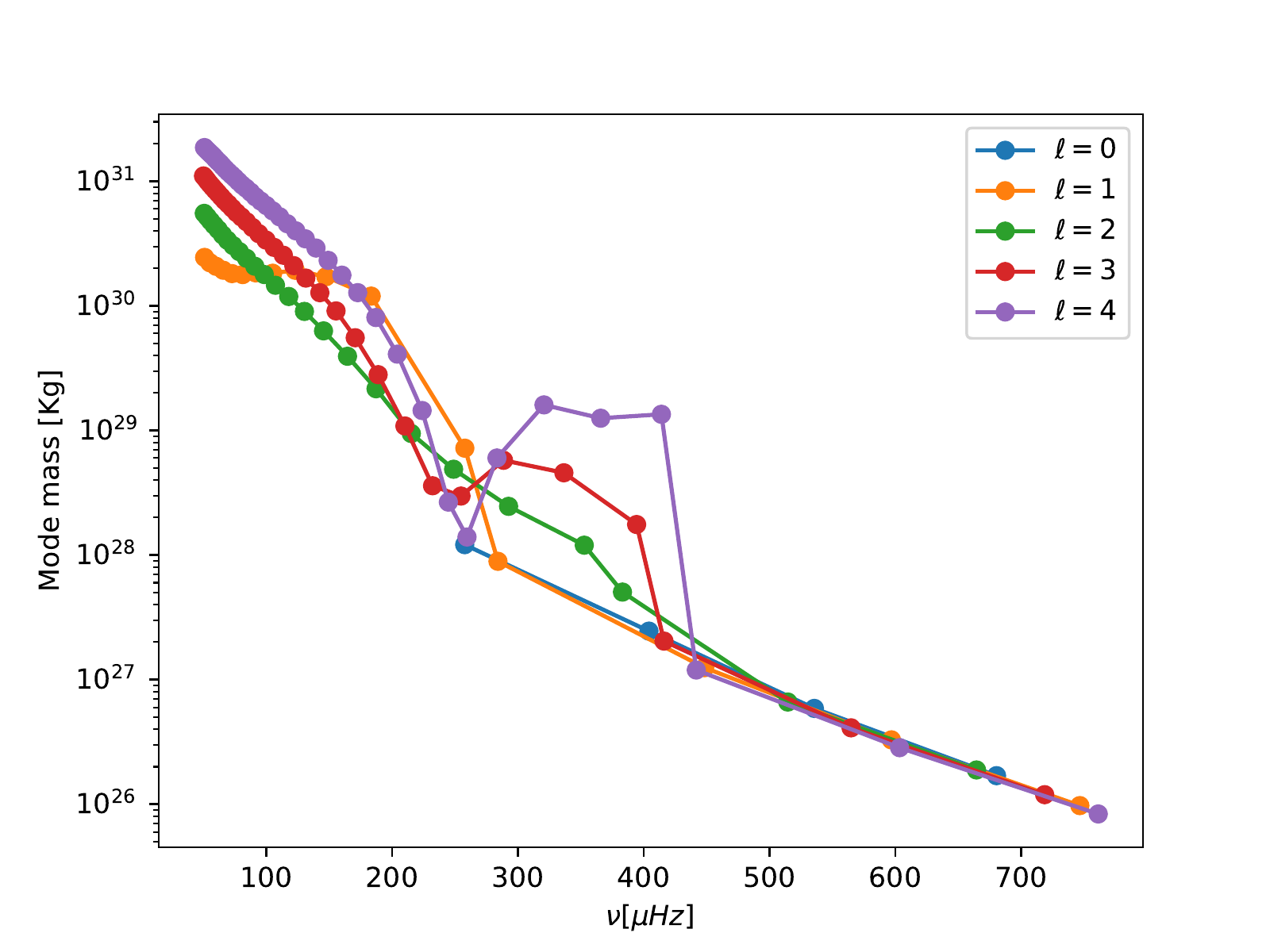}
		\caption{Mode masses (see Equation~\ref{mode_masses}) as a function of frequency computed using a seismic solar model as described by \citet{Buldgen2020}.}
			\label{fig_sun_mixed}
	\end{center}
\end{figure}

Estimating low-order $g$-mode amplitude is a challenging task because, as already mentioned in Section~\ref{solarlike} and \ref{driving_turb_mixed}, our knowledge of the physical mechanisms responsible for their damping is very approximative, to say the least. An immediate approach to overcome these difficulties would then be to extrapolate low-order $p$-mode amplitudes to the frequency range of the  closest low-order g-mode. However, these modes are actually mixed modes and as such deserve a particular treatment. In this section, we propose to do so by using the properties of mixed modes in a similar fashion as is done in evolved stars (see Section~\ref{driving_turb_mixed}). In the Sun, mixed-mode frequencies are found at the edge of the $p$- and $g$-mode frequency ranges, i.e. typically between $\nu \approx 260\,\mu$Hz and $\nu \approx 420\,\mu$Hz. Figure~\ref{fig_sun_mixed} displays the mode masses of solar modes, obtained by using a seismic solar model computed by \citet{Buldgen2020} (which agrees to a level of 0.1\,\% with seismic inversions of the solar structure) together with the LOSC oscillation code \citep{Scuflaire2008}. They are  defined as
\begin{equation}
\label{mode_masses}
 \mathcal{M} = \frac{1}{\vert
 {\bm \xi}
 (r=\mathrm{R}_{\odot}) \vert^2} \int_0^{\mathrm{M}_\odot} \vert
 {\bm \xi}
 \vert^2 {\rm d}m \, , 
\end{equation} 
where
$\bm{\xi}$
is the eigen-displacement vector, R$_\odot$ and M$_\odot$ are the solar radius and mass, respectively.
The sudden increase of the mode masses in Figure~\ref{fig_sun_mixed} between $\approx 280\,\mu$Hz and $420\,\mu$Hz is a signature of mixed modes for $\ell \ge 2$, which have a non-negligible amplitude in the inner layers of the Sun compared to radial and dipolar modes that behave as pure acoustic modes trapped in the outer envelope. 

Applying the same ideas as has been performed for evolved stars is therefore possible, with some adjustments. To do so, we first consider that at similar frequencies and with similar mode shapes in the super-adiabatic layers, the work done by the driving source on the modes is the same \citep[see][for details]{Dupret2009,Grosjean2014}. In addition, it is assumed that the radiative damping experienced by solar mixed modes in the inner radiative region can be neglected compared to the  damping located at the surface (as for $p$-modes). This assumption is reasonably well justified in the considered frequency range by the results of \citet{Dupret2009} and \citet{Grosjean2014}, who demonstrated that the radiative damping becomes dominant only for red giant stars in the vertical branch because of the important values of the buoyancy frequency compared to what is found in the Sun. In that case, the ratio of amplitudes between a mixed mode and a $p$-mode of similar frequency can be written \citep[][]{Dupret2009,Grosjean2014,Belkacem2019b}
\begin{equation}
\label{eq_ratio_amplitudes}
\frac{A_{\rm m}^2}{A_{\rm ref}^2} = \frac{\mathcal{M}_{\rm ref}}{\mathcal{M}_{\rm m}} \; , 
\end{equation}
where $\mathcal{M}_{\rm m}$, $\mathcal{M}_{\rm ref}$ are the mode masses (i.e. the normalized inertias) for the mixed mode and the reference $p$-like mode, respectively. $A_{\rm m}$ is the amplitude of the considered mixed modes and $A_{\rm ref}$ the amplitude of the reference mode which is assumed to be mainly of $p$-type. 

\begin{figure}[!ht]
	\begin{center}
		\includegraphics[width=0.9\textwidth,clip=]{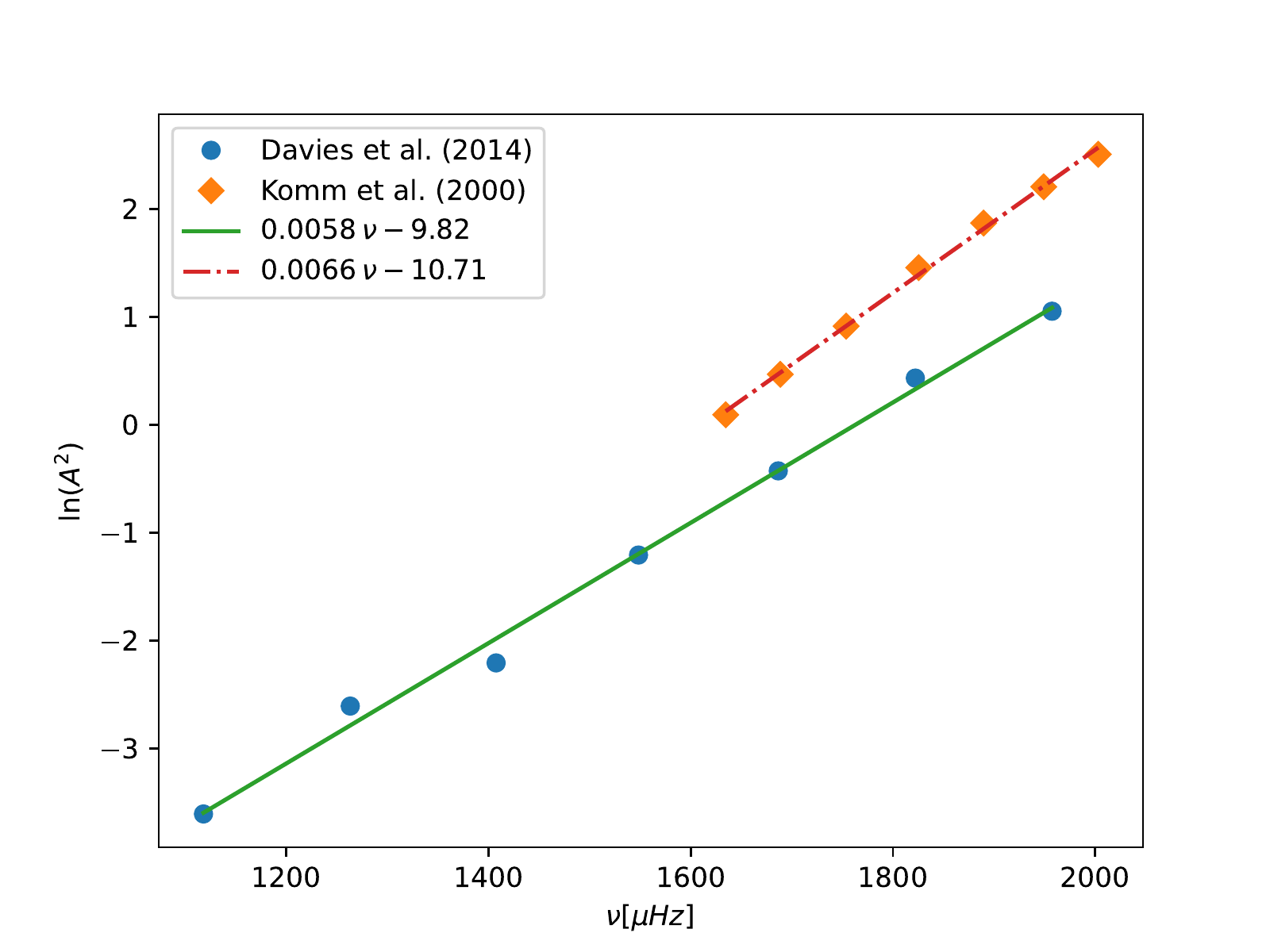}
		\caption{Natural logarithm of squared amplitudes for low-frequency modes as inferred by \citet{Komm2000} using the GONG (Global Oscillation Network Group) network and \citet{Davies2014} using the BiSON (Birmingham Solar Oscillations Network). The best fit have been obtained by minimizing the squared error. 
		}
	\label{amplitude_solar_mixed_pmodes}
	\end{center}
\end{figure}

While the objective is to estimate $A_{\rm m}$, one needs to determine $\mathcal{M}_{\rm ref},\mathcal{M}_{\rm m}$, and $A_{\rm ref}$ . To that end, we note that the reference modes and the mixed modes must have similar frequencies to ensure that, except for a normalization factor, they are affected by surface driving and damping in the same way. $\mathcal{M}_{\rm ref},\mathcal{M}_{\rm m}$ are therefore computed using the seismic solar model computed by \citet{Buldgen2020} together with the LOSC oscillation code \citep{Scuflaire2008} for radial modes and interpolated at the mixed mode frequencies. For $A_{\rm ref}$, we take advantage of the fact that low-frequency mode amplitudes (more precisely the natural logarithm of the squared amplitudes) linearly scale with the frequency, as shown by Figure~\ref{amplitude_solar_mixed_pmodes} \citep[see also][]{Komm2000,Salabert2009,Davies2014}. Therefore, we extrapolate the $p$-mode amplitudes as inferred by \citet{Komm2000} to the mixed-mode frequencies. In addition, due to instrumental effects and the low signal-to-noise ratio, inferences of low-frequency $p$-mode amplitudes are associated to non-negligible uncertainties. To assess its influence on mixed-mode amplitudes, we also use the amplitudes obtained by \citet{Davies2014} that provides slightly different scaling compared to \citet{Komm2000}, as shown by Figure~\ref{amplitude_solar_mixed_pmodes}. By doing so, it is assumed that the scaling holds until very low frequencies. We then stress that this is far from being a solid assumption and that the result must be considered as a rough order of magnitude estimate only. 

\begin{figure}[!ht]
	\begin{center}
		\includegraphics[width=0.9\textwidth,clip=]{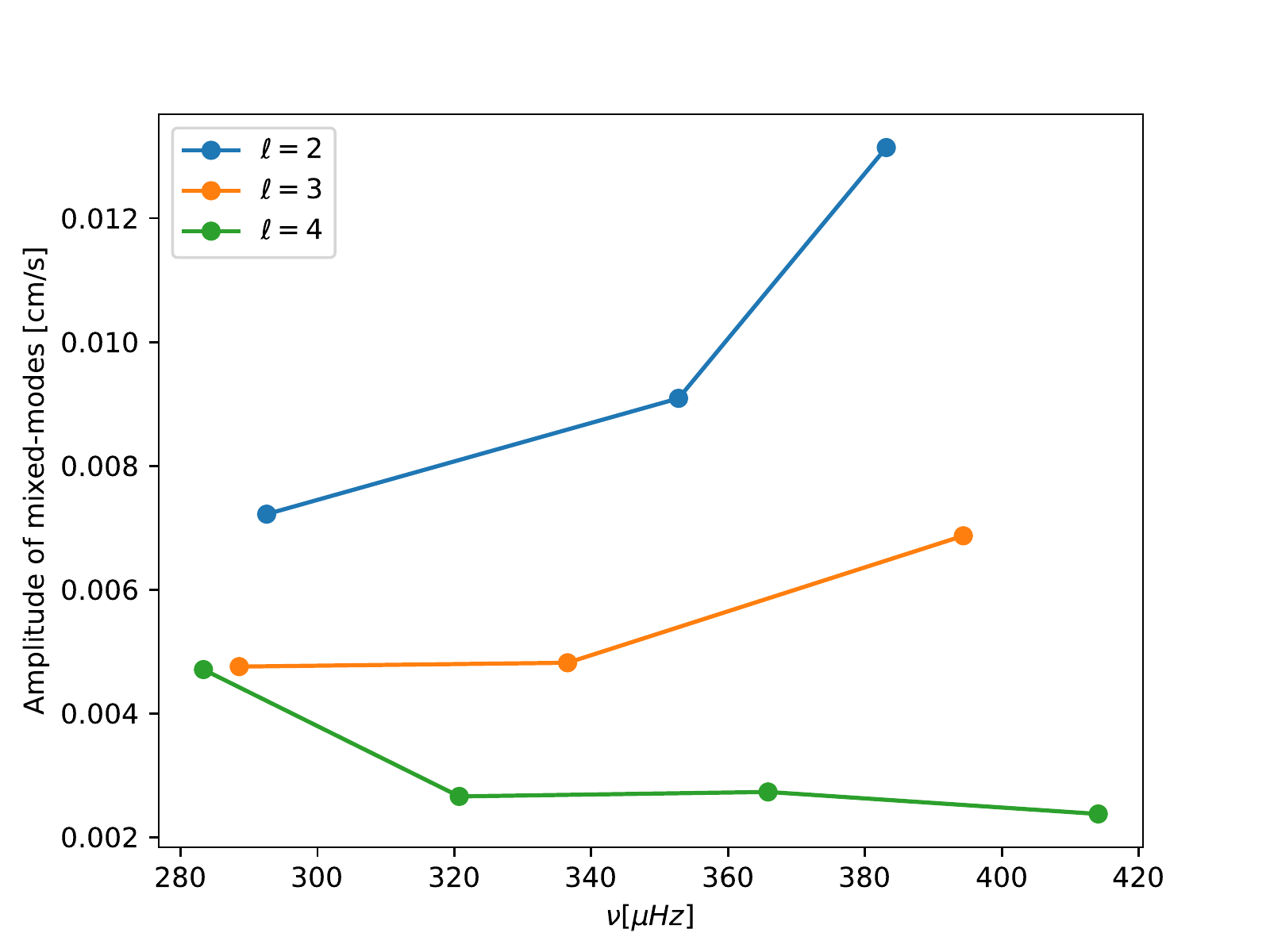}
		\caption{Amplitudes of $\ell=3$ and $\ell=4$ solar mixed modes versus mode frequencies. The amplitudes have been computed as described in Section~\ref{analogy_g}. Note that, radial and dipolar modes in this frequency range being essentially of acoustic nature, we did not consider them. 
		}
	\label{amplitude_solar_mixed}
	\end{center}
\end{figure}

The result is displayed in Figure~\ref{amplitude_solar_mixed} for mixed-modes of angular degrees $\ell=2,3,$ and $4$. It turns out that the amplitudes are relatively small (well below a few millimeters per second) and, as one can guess from \eq{eq_ratio_amplitudes}, the larger amplitudes are found for quadripolar modes. Note that radial and dipolar modes, in the same frequency range, certainly have higher amplitudes if one extrapolated the amplitudes observed by \citet{Komm2000} but these modes being essentially of acoustic nature they hardly bear information on the inner-most layers of the Sun. It also turns out that using the amplitudes inferred by \citet{Davies2014} rather than \citet{Komm2000} only modifies the amplitudes by about $30\,\%$ and it thus does not affect the main conclusion of this simple estimate.

\section{The Solar Gravity Modes: Driving by Penetrative Convection}
\label{penetrative_conv}

Excitation by turbulent convection is not the only way to generate gravity modes. Penetrative convection is also thought to be an efficient mechanism to excite internal waves as has been known for many years for geophysical flows \citep[][]{Stull1976}. In the Sun, turbulent plumes are created at the upper boundary of the convection zone,
where radiative cooling becomes dominant and where the flow reaches the stable atmosphere. In this region, the updrafts become cooler than their environment and stop their ascent. This cool flow is then denser than its environment and triggers the formation of turbulent descending plumes \citep[e.g.][]{Stein98}. When plumes fall down through the convection zone, they entrain the surrounding flow at their edge.
It is the {\it entrainment hypothesis}, first introduced by G.I. Taylor and supported by observations in geophysics \citep[for a review see][]{Turner86}. This leads to the formation of large-scale downwelling turbulent structures that reach the stably stratified radiative zone below and that penetrate over some distance releasing their kinetic energy into internal waves and presumably into modes. 
In contrast with the driving by turbulent pressure, an analogy with the excitation and damping of the observed solar $p$-modes is not possible. Current theoretical estimates of the excitation by penetrative convection therefore rely on either numerical experiments based on hydrodynamical simulations or semi-analytical models. In this section, we give a brief overview of the current state-of-the-art about this complex question.

\subsection{Help of Hydrodynamical Simulations: Useful but Limited Insights}
\label{help_simu3D}

\citet{Andersen96} was one of the first to propose a theoretical estimation of solar $g$-mode amplitudes generated by penetrative convection using 2D Cartesian numerical simulations of the solar convective region. \citet{Andersen96} studied how gravity waves generated at the base of the convective zone, by ad-hoc oscillatory perturbations mimicking penetrating turbulent plumes, can tunnel toward the surface of the Sun. Doing so, he could compute the associated attenuation factors that measure the exponential decrease of the $g$-mode amplitudes in the convective region. Using heuristic energetic considerations and these attenuation factors, they extrapolated $g$-mode amplitudes at the Sun surface from $0.01$ to $5\,\textrm{mm.s}^{-1}$. Nevertheless, those results are based on rather crude assumptions. For instance, they largely depend on the chosen number of modes over which the total oscillation energy injected from convective plumes is distributed. More importantly, the reasoning does not properly consider damping and trapping processes in the underlying radiative zone. 

More recently, \cite{Boris05} also investigated the excitation of $g$-modes by penetrative convection using 2D Cartesian simulations, but in a more consistent way. In this work, the authors modeled a star as composed of three polytropic layers: a bottom stable radiative layer, an intermediate convective layer, and a very thin stable cooling upper layer. Their simulation exhibits well-developed downward plumes that can perturb the thermal stratification at the base of the convective region, generating progressive gravity waves in the underlying radiative layers. After a large enough computing time, $g$-modes subsequently emerge. The authors then could study the relation between plume penetration and $g$-mode amplitudes based on the projection of the observed velocity field onto the theoretical $g$-mode eigenfunction basis of the 2D simulated box. Using this technique, they could overcome the difficulty to disentangle the motions associated with $g$-modes and those associated with turbulent convection throughout the simulation for different time spans. Their results first highlight the random properties of the excitation process. They showed in particular that $g$-modes generated by different penetrating plumes negatively interfere on average. These destructive interferences occur over timescales corresponding to about twice the modal period, demonstrating the correlation between the occurrence rate of penetration events and the $g$-mode frequency range that is generated. Second, they showed that up to $40\,\%$ of the total kinetic energy of the simulated box can reside in $g$-modes. This result was very promising and supported penetrative convection as an efficient mechanism to generate $g$-modes in stars. The simulation setup was however too simple to allow to infer quantitatively the amplitude of $g$-modes in the Sun ( polytropic stratification, Cartesian geometry) and called for additional investigations.  

Pushing realism one step further, \cite{Rogers05} performed 2D numerical simulations of solar equatorial slices with a realistic hydrostatic stratification. A striking result of this study is that using quasi-linear simulations (i.e. neglecting the non-linear terms in the radiative zone only), $g$-modes can be excited by penetrative convection and can survive over time. In contrast, fully non-linear simulations do not exhibit $g$-modes but only low-frequency progressive waves. Non-linear wave--wave interaction is thought to be responsible for this result. A similar mechanism was proposed by \cite{Kumar89} to explain $p$-mode damping. These authors indeed demonstrated that the dominant non-linear effect can couple three waves, two trapped $p$-modes whose total energy is drained by a third propagative wave. Nevertheless, to ensure numerical stability and reasonable computing times, such simulations with a realistic solar thermal stratification actually require much larger thermal diffusivities than in the Sun. In this study, the thermal diffusivities were increased by a factor of about 10$^5$, which resulted in an overestimate of the solar convective flux by the same factor \citep{Rempel04}. Hence, one can expect that the generated wave energy flux is also overestimated, leading to very large mode amplitudes. The claim of \cite{Rogers05} that $g$-modes generated by penetrative convection are likely to not exist in the Sun because of non-linear effects is thus to be taken with caution since non-linearity is very likely the result of the artificially enhanced diffusivities considered in their simulations.

Extending these previous studies to 3D spherical geometry is nowadays possible. Such simulations were for instance performed by \cite{Alvan2014}. This study also considered a realistic solar thermal stratification as the  initial background state, still at the expense of a thermal diffusivity ineluctably enhanced by a factor of about $10^5$ compared to the solar interior. In contrast with the work of \cite{Rogers05}, the authors observed that penetrative plumes are able to trigger the formation of $g$-modes even in the fully non-linear case. Actually, in this work, the nuclear energy production rate at the center was reduced in such way as to conserve a solar luminosity at the surface of the simulation. This condition in addition to a large thermal diffusivity resulted in much slower convective plumes and thus lower $g$-mode amplitudes than by \cite{Rogers05}, hence reducing modal non-linear effects in their simulations. \cite{Alvan2014} then could estimate the amplitude of $g$-modes at the top of the radiative zone. They predicted that the highest amplitude for $\ell=1-3$ in the frequency range $\nu\approx 50 - 400\,\mu$Hz goes from 10$^{-11}$ to 10$^{-5}\,\textrm{cm.s}^{-1}$ as the thermal diffusivity is reduced by three orders of magnitude in the radiative zone and the Reynolds number in the convective region is increased by a factor of only four. This increasing trend is not surprising. Indeed, on the one hand, $g$-mode amplitudes are inversely proportional to the mode damping, which is proportional to the radiative diffusivity in the frequency regime observed in the simulation. On the other hand, increasing the Reynolds number in the convective region is also expected to increase the excitation of $g$-mode as convective flows become more turbulent and vigorous. Extrapolating this promising trend towards much higher Reynolds numbers and much lower thermal diffusivities expected in the solar regime is quite tempting and would lead to very high $g$-mode amplitudes, certainly much higher than the current detection threshold. However, these results still need to be taken with caution.
First, it is not clear how much the mean thermal structure of these simulations deviates from the initial solar-structure model after a thermal relaxation timescale. Indeed, as the total luminosity is kept to the solar-luminosity value, the thermal gradient needs to adjust to the very large value of the diffusivity before reaching a new thermal equilibrium. This could significantly modify the Brunt--Väisälä frequency profile in the simulated bulk and thus the damping, driving and propagation of the $g$-modes, hence at odds with the initial assumption of a realistic solar stratification. Second, the dependence of the mode damping and driving with the thermal diffusivity and the turbulence level in the convective region, as well as their interplay, also needed to be clarified.

Recently, \cite{Lesaux22} attacked this question in more detail using 2D simulations of solar meridional slices. The authors considered a realistic stratification, varying the value of the thermal diffusivity and thus the total luminosity by several orders of magnitude. Following  \cite{Baraffe21}, the authors took special care to choose an initial thermal background state that is close to the solar one and close to the thermally relaxed state of the simulations to avoid contaminating their study by uncontrolled structural modifications. All their set of simulations clearly showed that low-frequency progressive gravity waves and high-frequency $g$-modes are simultaneously excited by both turbulent pressure and penetrative plumes. Their results confirmed that the convective velocities and thus the inverse of the typical convective turnover timescale scales as the luminosity to the power $1/3$. However, this is not the case of the $g$-mode amplitudes. As the stellar luminosity is enhanced, the convective energy transferred into $g$-modes is distributed over higher frequencies as the typical convective turnover timescale decreases. The excited range of spherical degree $\ell$ is also modified in a different way. This change in the $g$-mode driving due to the enhancement in the thermal diffusivity, while keeping a quasi constant solar stratification, modifies the mode damping and inertia in the radiative resonant cavity in a non-linear way, making an extrapolation towards the solar case difficult. Keeping in mind that the current simulations stand just above the convection threshold and are far from turbulent regimes expected in stars, the authors then concluded that we have to be very careful when interpreting the studies of $g$-modes based on hydrodynamical simulations with artificially enhanced luminosity.

Overall, numerical hydrodynamical simulations represent a useful experimental bench to study the excitation of the solar $g$-modes by penetrative convection and can provide useful qualitative insights. However, they still remain limited and a quantitative estimate of the $g$-mode amplitudes seems today out of reach. To go further, a complementary and necessary approach therefore consists in estimating the $g$-mode amplitude by semi-analytical models.
 
\subsection{Quantitative Hints From Semianalytical Models}
\label{semi_analytics}

A number of semi-analytical models exist to estimate the generation of progressive gravity waves at the interface between the convective envelope and the radiative core of the Sun, but most of these developments consider that the driving is controlled by turbulent convective rolls, that is, using a picture very similar to the excitation by turbulent pressure \citep[][]{Press81,Lecoanet2013}. To complete these prescriptions, \cite{Pincon16} proposed a new estimate considering penetrative convection as the driving mechanism, which was later adapted to the question of the excitation of the stationary solar $g$-modes by \cite{Pincon2021}. 

In this latter work, the authors modeled the excitation process using four main assumptions:
\begin{enumerate} 
\item The source term in the linear oscillation equations corresponds to the dynamical ram pressure exerted by an ensemble of incoherent plumes radially penetrating at the base of the convective region and uniformly distributed over the sphere. 
\item There is no feedback from the $g$-modes to the plume behavior. 
\item The limiting case of a very large plume P\' eclet number at the base of the convective zone is assumed. This means that the density contrast between the plumes and the surroundings is high at the top of the radiative region so that the buoyancy braking of the plumes is very strong in the penetration zone, as expected in the Sun. 
\item The frequency range considered lies between 10\,$\mu$Hz and 100\,$\mu$Hz. Indeed, in this frequency range, the $g$-mode wave function can first be easily deduced using the JWKB asymptotic method within the short-wavelength assumption; second, the $g$-mode damping is dominated by radiative diffusion and the mode lifetime is much larger than the mode period, which enables us to treat simultaneously and consistently the driving and the damping using a two-time method. 
\end{enumerate}
Finally, \cite{Pincon2021} expressed the mean energy of a $g$-mode with radial order $n$, angular degree $\ell$, and angular frequency $\omega_{n\ell}$, in the compact analytical form 
\begin{equation}
\langle E_{n\ell } \rangle \approx \frac{1}{\pi n} \frac{  \left[{L_{\rm p}}~ F_{{\rm d},\ell} ~\mathcal{H}_\ell ~\mathcal{C}_{n\ell}\right]}{ \eta_{n\ell}} \; ,
\label{mean energy}
\end{equation} 
where \smash{${L_{\rm p}}$} is the mean plume kinetic luminosity at the base of the convective zone, $F_{{\rm d},\ell}$ is the plume Froude number, which measures the efficiency of the energy transfer from the plumes to the $g$-modes, and $\eta_{n\ell}$ is the damping rate per unit of time. Finally, the $\mathcal{H}_\ell$ and $\mathcal{C}_{n\ell}$ terms measure the horizontal and temporal correlations between the plumes and the modes.

Using a 1D solar-structure model, they could estimate the damping rate. Besides, they estimated the plume width and velocity using the semi-analytical turbulent model of plumes developed by \cite{Rieutord95}, providing an estimate of ${L}_{\rm p}$, $\mathcal{H}_\ell$, and $F_{{\rm d},\ell}$. The most critical point of their work was then to model the temporal-correlation term. Indeed, the authors showed that the $g$-mode energy is very sensitive to the plume temporal-evolution profile inside the penetration zone. The characteristic plume lifetime was first chosen around the turnover timescale of the convective eddies above the base of the convective zone, as predicted by the mixing length theory in the 1D solar model. Such a choice appears judicious as it is supported by recent hydrodynamical simulations \cite[][]{Lesaux22}.
The hard part was then to model the shape of the plume temporal-evolution profile, which still remains poorly constrained. Considering both limiting cases of a Gaussian and exponential evolution over time, they demonstrated that the temporal correlation is very much larger in the exponential than in the Gaussian case. Using appropriate visibility factors, the authors could then compute the apparent radial mode velocities at the solar surface based on their predictions of the mean mode energy. The mode amplitudes turned out to be maximum for the $\ell=1$ $g$-modes and to increase with mode frequency. For $\nu\approx  100\,\mu$Hz, the $g$-mode amplitudes can reach $0.05\,\textrm{cm.s}^{-1}$ in the exponential case, which is still one order of magnitude lower than the current GOLF detection threshold and the estimate considering the turbulent Reynolds stress as the driving mechanism by \cite{Belkacem2009}. Considering uncertainties in the plume parameters, reasonable variations in their values in the most plausible favorable case (i.e. the plume radius and occurrence rate are multiplied by a factor of about two) can lead to an increase of the apparent mode radial velocity to $0.4\,\textrm{cm.s}^{-1}$ at $\nu\approx 100\,\mu$Hz, which still requires at least 25 years of GOLF observations to be detected (see Figure~\ref{bilan_amplitudes}).

\section{Concluding Remarks and Prospects}
\label{s:conclusion}

\begin{figure}[!ht]
	\begin{center}
		\includegraphics[width=12cm]{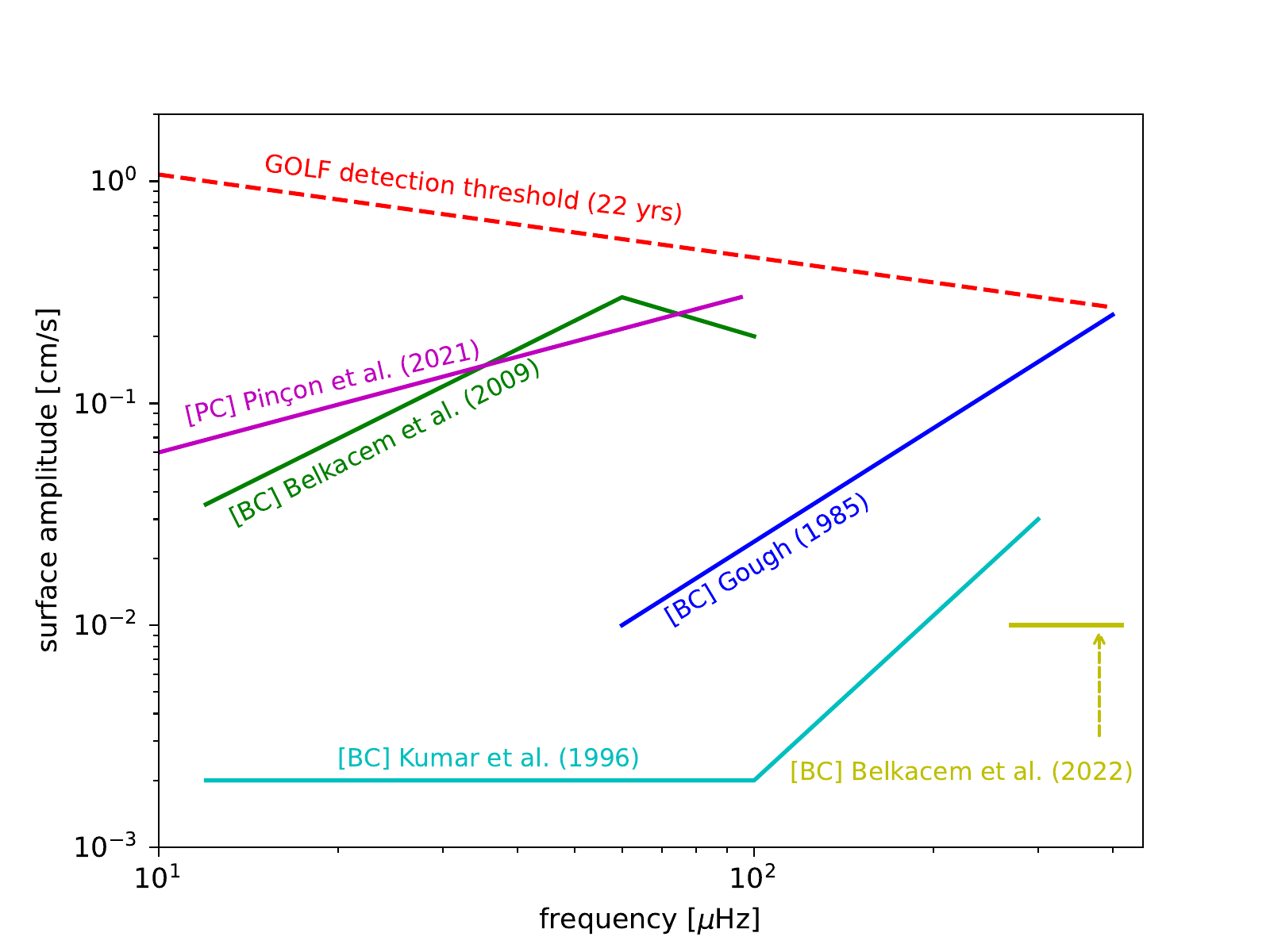}
		\caption{Schematic mode surface velocity as function of frequency. Theoretical computations by \citet{Gough85,Kumar96,Belkacem2009,Pincon2021} as well as the calculation described in Section~\ref{analogy_g} (labelled as Belkacem et al. 2022) are provided in \textit{solid lines} while the GOLF observational threshold associated with 22 years of
observation is in \textit{dashed line} and is computed as described by \citet{Pincon2021}. The theoretical computations are labeled with either BC (for Bulk Convection, which denotes that the calculations have been done assuming that the main source of driving is turbulent convection inside the convective region) or PC  (for Penetrative Convection, which denotes that the calculations have been done assuming that the main source of driving is due to turbulent plumes penetrating into the stably stratified region).
	}
	\label{bilan_amplitudes}
	\end{center}
\end{figure}

Detection of solar gravity modes, despite many false positives, is still a major objective in solar physics as it is a unique probe of the innermost layers. Indeed, determining the rotation of the solar core would provide an important anchoring point for the understanding of angular momentum processes in main-sequence solar-type stars, as well as providing important tests for the physical conditions of the solar core complementary to neutrino experiments. Because some review articles \citep[such as][]{Appourchaux2010} already provide in-depth discussions on the challenges from different perspectives (instrumental, data analysis, theoretical aspects, etc...), we focused our discussion on theoretical estimates of $g$-modes amplitudes by putting them in perspective regarding recent advances in asteroseismology. Indeed, thanks to space-based missions such as CoRoT \citep[e.g.][]{Baglin2006a,Baglin2006b} and {\it Kepler} \citep[e.g.][]{Borucki2010} and the detection of hundreds of main-sequence stars and thousands of evolved stars exhibiting solar-like oscillations \citep[e.g.][]{Chaplin2013}, our knowledge about the physics of normal modes in low-mass stars significantly improved and in particular regarding the detection and characterization of mixed-modes in evolved stars
\citep[e.g.][]{Dupret2009,Grosjean2014,Belkacem2019b}. 

In that context, it is worth revisiting previous and recent estimates of solar $g$-mode amplitudes and this was done in this article. 
This is summarized in Figure~\ref{bilan_amplitudes}, which provides most of the discussed estimates as well as the GOLF detection threshold after 22 years of continuous observations. Overall, these estimates, considering the excitation both by turbulent pressure and penetrative convection, suggest that the low-frequency solar gravity modes are currently at best at the limit of the detection  with the GOLF instrument. At higher frequencies, i.e. near the edge of the observed $p$-mode frequency range, the situation is more contrasted. While early theoretical calculations by \citet{Gough85} were optimistic about the probability of a positive detection, advances on the modeling of mode driving and damping by turbulent convection led us to revise this estimate. Indeed, in this article, using solar mixed modes and based on  methods developed for mixed-modes in evolved low-mass stars, we confirm this pessimistic view as shown by Figure~\ref{bilan_amplitudes}. Therefore, this confirms our current view that $g$-modes are more likely to be detected at low frequencies (in the asymptotic regime) while low-order $g$-modes appear to be out-of-reach using our current observational facilities. 

Nevertheless, one must keep in mind that estimating $g$-mode amplitudes is a very challenging task. Indeed, our knowledge of mode damping in the convective region remains quite limited and, even worse, our knowledge of the physical properties of the solar turbulent convection itself is only based on very simplified models (except for the surface layers) mainly based on the mixing-length theories. However, as discussed in this article, precise and accurate knowledge of the dynamical properties of the solar convective region is key for assessing solar $g$-mode amplitudes. Fortunately, some encouraging perspectives exist and they are two-fold: first, hydrodynamical and magnetohydrodynamical simulations (even if they are still far from the physical regimes encountered in the solar interior) are of great help for inferring the physical properties of the turbulent convective region and, given the increase of numerical capacities, their degree of realism is improving relatively rapidly. Second, the recent detection of solar inertial modes brings great promise to get a direct probe of the convective regions, even the deep convection zone, as was nicely demonstrated by \citet{Gizon2021}. 

\begin{acks}
The authors thank the reviewer for their useful comments that helped to improve the manuscript. 
\end{acks}
\begin{ethics}
\begin{conflict}
The authors declare that they have no conflicts of interest.
\end{conflict}
\end{ethics}
\begin{fundinginformation}
During this work, C. Pin\c{c}on  was financially supported by Sorbonne University (EMERGENCE project). G. Buldgen acknowledges funding from the SNF AMBIZIONE grant No 185805 (Seismic inversions and modelling of transport processes in stars).
\end{fundinginformation}

\bibliographystyle{spr-mp-sola}

\end{article} 

\end{document}